\documentclass[letterpaper, 10 pt, conference]{ieeeconf}  %
\IEEEoverridecommandlockouts  
\usepackage{amsmath}
\usepackage{amssymb}
\usepackage{graphicx}
\usepackage{amssymb,latexsym,amsfonts,amsmath}
\usepackage{mathrsfs}
\usepackage{graphicx}
\usepackage{eso-pic}
\usepackage{epsfig}
\usepackage{mathtools}
\usepackage{tikz}
\usepackage{mdwlist}
\usepackage{refcount}
\usepackage{comment}
\usepackage{caption}
\usepackage{subfig}
\let\labelindent\IEEElabelindent
\usepackage{enumitem}
\usepackage{keyval,trig}
\usepackage{amssymb,dsfont}
\usepackage{mathtools,mathrsfs}
\usepackage{float}
\usepackage{cite}
\usepackage[linesnumbered,ruled]{algorithm2e}

\usepackage{algorithm}
\usepackage{tikz}
\usepackage{xspace}
\usepackage[many]{tcolorbox}
\usepackage{caption}
\usetikzlibrary{calc}
\tcbuselibrary{skins}
\newtheorem{theorem}{Theorem}[section]

\newtheorem{problem}[theorem]{Problem}

\newtheorem{remark}[theorem]{Remark}

\newcommand{\always}{\Box}
\newcommand{\eventually}{\Diamond}
\newcommand{\until}{\mathbin{\sf U}}

\newtcolorbox{resp1}[1][]{%
	enhanced jigsaw,%
	colback=gray!5!white,%
	colframe=gray!80!black,%
	size=small,%
	boxrule=1pt,%
	halign title=flush center,%
	coltitle=black,%
	breakable,%
	drop shadow=black!50!white,%
	attach boxed title to top left={xshift=1cm,yshift=-\tcboxedtitleheight/2,yshifttext=-\tcboxedtitleheight/2},%
	minipage boxed title=3cm,%
	boxed title style={%
		colback=white,%
		size=fbox,%
		boxrule=1pt,%
		boxsep=2pt,%
		underlay={%
			\coordinate (dotA) at ($(interior.west) + (-0.5pt,0)$);
			\coordinate (dotB) at ($(interior.east) + (0.5pt,0)$);
			\begin{scope}[gray!80!black]
				\fill (dotA) circle (2pt);
				\fill (dotB) circle (2pt);
			\end{scope}
		}%
	},%
	#1%
}

\usepackage[colorlinks=true, citecolor=blue, linkcolor=blue, final]{hyperref}
\xdefinecolor{MATLABblue}{rgb}{0.0,0.45,.74}
\xdefinecolor{MATLABred}{rgb}{0.85,0.33,.1}

\newcommand{\ea}{{ et al. }}

\newcommand{\st}{{\rm s.t.}}

\IEEEoverridecommandlockouts

\newcommand{\Negar}{\textcolor{blue}}
\newtcolorbox{resp}[1][]{%
	enhanced jigsaw,%
	colback=gray!5!white,%
	colframe=gray!80!black,%
	size=small,%
	boxrule=1pt,%
	halign title=flush center,%
	coltitle=black,%
	breakable,%
	drop shadow=black!50!white,%
	attach boxed title to top left={xshift=1cm,yshift=-\tcboxedtitleheight/2,yshifttext=-\tcboxedtitleheight/2},%
	minipage boxed title=3cm,%
	boxed title style={%
		colback=white,%
		size=fbox,%
		boxrule=1pt,%
		boxsep=2pt,%
		underlay={%
			\coordinate (dotA) at ($(interior.west) + (-0.5pt,0)$);
			\coordinate (dotB) at ($(interior.east) + (0.5pt,0)$);
			\begin{scope}[gray!80!black]
				\fill (dotA) circle (2pt);
				\fill (dotB) circle (2pt);
			\end{scope}
		}%
	},%
	#1%
}
\title{\LARGE \bf
Logic-based Resilience Computation of Power Systems Against Frequency Requirements
}
\author{Negar Monir$^{1}$, Mahdieh S. Sadabadi$^{2}$, and Sadegh Soudjani$^{3}$
\thanks{*The research of N. Monir is supported by the EPSRC EP/W524700/1 and Newcastle University Global Scholarship. The research of S. Soudjani is supported by the following grants: EIC 101070802 and ERC 101089047.}
\thanks{$^{1}$N. Monir is with the School of Computing, Newcastle University, Newcastle upon Tyne, United Kingdom
        {\tt\small (s.seyedmonir2@ncl.ac.uk)}}%
\thanks{$^{2}$M. S. Sadabadi is with the Department of Electrical and Electronic Engineering, The University of Manchester, Manchester, United Kingdom
        {\tt\small (mahdieh.sadabadi@manchester.ac.uk)}}%
\thanks{$^{3}$S. Soudjani is with Max Planck Institute for Software Systems, Germany
        {\tt\small (sadegh@mpi-sws.org)}}%
}
\begin{document}
\maketitle
\thispagestyle{empty}
\pagestyle{empty}
\begin{abstract}

Incorporating renewable energy sources into modern power grids has significantly decreased system inertia, which has raised concerns about power system vulnerability to disturbances and frequency instability. The conventional methods for evaluating transient stability by bounding frequency deviations are often conservative and may not accurately reflect real-world conditions and operational constraints. To address this, we propose a framework for assessing the resilience of power systems against disturbances while adhering to realistic operational frequency constraints. Our approach leverages the Lur’e system representation of power system dynamics and Signal Temporal Logic (STL) to capture the essential frequency response requirements set by grid operators. This enables us to translate frequency constraints into formal robustness semantics. We then formulate an optimization problem to identify the maximum disturbance that the system can withstand without violating these constraints. The resulting optimization is translated into a scenario optimization while addressing the uncertainty in the obtained solution. The proposed methodology has been simulated on the Single Machine Infinite Bus case study and 9-Bus IEEE benchmark system, demonstrating its effectiveness in assessing resilience across various operating conditions and delivering promising results.
\end{abstract}
\section{INTRODUCTION}
Global energy systems are currently undergoing a significant transformation in response to the urgent challenges posed by climate change, environmental sustainability, and the need to ensure the reliability and resilience of energy systems. This shift involves a growing emphasis on renewable energy sources, such as wind and solar, to reduce reliance on fossil fuels and minimize their negative environmental effects. Power systems, as a critical component of energy infrastructure, play a crucial role in both national and international stability, resilience, and economic welfare \cite{HAESALHELOU2021107477, OLASOJI2024e02357}. While these systems have traditionally operated reliably for many years, with occasional blackouts and outages, the increasing integration of renewable energy sources has introduced new challenges, particularly in terms of the stability and resilience of modernized power systems. A primary concern in these systems is the need to maintain power system frequency within mandatory operational limits set by grid operators. Adhering to these frequency constraints is essential for ensuring the overall resilience and stability of the power grid \cite{kundur2007power, vittal2019power, teng2015benefits, lee2019robustness,wooding2020formal}.

In the field of power systems, various methods have been developed to evaluate transient stability under different operational constraints. Traditionally, transient stability analysis focused on determining if a power system could maintain its synchrony after major disturbances, such as faults or load shedding, by assessing the system's ability to restore frequencies and rotor angles to their steady-state values \cite{chiang2011direct, miller2011freq, wilches2016fund}. However, the increasing complexity of power systems due to the integration of renewable energy sources has necessitated more sophisticated assessment methods. These methods can be broadly categorized into three groups. The first group involves numerical simulations, which model disturbances and operational conditions in detail but often require significant computational resources \cite{dong2012numeric,milano2013systematic,papado2017probabilistic}. The second group uses reachability analysis to provide tight time-dependent bounds on system states by modeling dynamics with linear approximations or differential-algebraic equations. Although effective, these methods involve solving optimization problems iteratively, making them computationally intensive \cite{chen2012method,althoff2014,choi2017propagating,lee2017robust,zhang2019characterizing}. The third group leverages input-to-state stability (ISS) analysis, offering a mathematically rigorous approach but often encountering difficulties in constructing appropriate Lyapunov functions \cite{aolaritei2018robustness,WEITENBERG20181}.

The paper by Lee\ea \cite{lee2019robustness} introduces a new framework for assessing the transient stability of power systems. They use input-output stability analysis for nonlinear Lur'e systems representation of the power systems' dynamical model. The proposed approach in \cite{lee2019robustness} focuses on evaluating the grid's robustness against disturbances by characterizing the calculation of the largest disturbances as an optimization problem. This allows for a mathematically rigorous assessment of transient stability under frequency constraints in power systems. The proposed method efficiently establishes certificates of robustness and enables real-time validation against a class of magnitude-bounded disturbances. However, using an infinite norm to define a magnitude-bounded frequency is conservative, and may not fully capture the time-varying nature of frequency requirements and operational frequency constraints in modern power grids. This limitation points to the need for more refined models that incorporate temporal logic specifications to define acceptable frequency ranges and operational constraints over time.

This paper seeks to develop a new method for assessing the resilience of power systems to disturbances while meeting frequency constraints set by grid operators, such as the National Grid of the United Kingdom (UK), the European Grid, the German Grid, and others.Our work aims to rigorously define frequency requirements and identify external disturbance levels under which the power system maintains frequency constraints. Unlike prior approaches that consider solely bounded frequency response, our method relies on  a less conservative approach by considering frequency characteristics. To achieve this, we discretize the Lur'e representation of the power systems' dynamic model \cite{khalil2002nonlinear}. We then establish frequency constraints for the power system based on the UK National Grid Code \cite{gridcode2023}, as a case study. These constraints are articulated as temporal logic specifications using the robustness semantics of Signal Temporal Logic (STL) \cite{oded2004monitoring, jin2013mining}. 
STL robustness and temporal logic specifications have gained attraction in analyzing power systems and smart grids applications. These methods provide a framework for specifying and evaluating system performance, enabling effective monitoring and control \cite{taousser2020model, ma2020sastl}. 
By translating the frequency constraints into STL robustness metrics, we can quantify the resilience of the power system against various disturbances.

We formulate a {nonconvex and non-smooth} optimization problem based on the system model and the STL-based frequency constraints to determine the largest disturbances that the system can handle without breaching the STL-based frequency requirements. To address the uncertainty inherent in the power system dynamics, {modeled as an uncertainty in the nonconvex optimization problem}, we propose using the scenario optimization approach \cite{campi2018wait, campi2018general, garatti2024non}. The scenario optimization method has gained popularity in solving robust optimization problems \cite{hu2021optimal,liu2023two,dietrich2024nonconvex, kordabad2024distributionally}. It allows us to translate the original optimization problem into a computationally tractable form, where we can identify the maximum disturbance the system can withstand while satisfying the STL-based frequency constraints. This approach provides a comprehensive assessment of the power system's resilience, going beyond the conservative bounds often produced by conventional stability analysis methods. The effectiveness of the proposed resilience framework is evaluated through analysis and simulation using IEEE 9-bus benchmark system\cite{IEEE9bus}.

The paper is organized as follows. The preliminaries and problem formulation are presented in Section \ref{sec: prele} and Section \ref{sec: prob form}, respectively. Section \ref{sec: large dist}  is devoted to solving the optimization problem and calculating the largest disturbance. Simulation results are illustrated in Section \ref{sec: sim res}. Finally, Section \ref{sec: concl} concludes this paper.
\section{PRELIMINARIES} \label{sec: prele}
In this section, we present the preliminaries required for our problem.
The power grid can be modeled as an undirected graph $\mathcal{G}(\mathcal{N}, \mathcal{E})$, where $\mathcal{N} = \{1, 2, \ldots, n\}$ represents the set of buses, and $\mathcal{E} \subseteq \mathcal{N} \times \mathcal{N}$ indicates the set of transmission lines that connect these buses. The total number of transmission lines is denoted as $\ell = |\mathcal{E}|$. Generation units are indexed by $\mathcal{G}_G = \{1, \ldots, m\}$, while loads are indexed by $\mathcal{G}_L = \{m + 1, \ldots, m + n\}$. The graph's incidence matrix is represented by {$E \in \mathbb{R}^{n \times \ell}$}. Furthermore, let $\mathbf{0}$ and $I$ be the zero matrix and the identity matrix with appropriate dimensions. The spectral radius of a matrix {$A \in \mathbb{R}^{n \times n}$} is denoted as $\varrho(A)$.
\subsection{Power System Model}
The following second-order swing equation, which retains the structural characteristics of the power system, serves as a model for analyzing power system dynamics.
\begin{align}
 M_k \ddot{\delta}_k+D_k \dot{\delta}_k+\sum_{(k, j) \in \mathcal{E}} \phi_{k j} \sin \left(\delta_{k j}\right)&=p_k, \quad \forall k \in \mathcal{G}_G \nonumber \\
D_k \dot{\delta}_k+\sum_{(k, j) \in \mathcal{E}} \phi_{k j} \sin \left(\delta_{k j}\right)&=P_{L, k}, \quad \forall k \in \mathcal{G}_L \label{eq: 2ndO Power Sys Dy}
\end{align}
where $M_{k}$ and $D_{k}$ represents the inertia and damping coefficients of the generation unit $k$, respectively. In \eqref{eq: 2ndO Power Sys Dy}, the mechanical power input at generation unit $k$ is represented by $p_k$, while $P_{L,k}$ represents the power demand at bus load $k$. The term $\phi_{k j}=b_{k j} V_{k} V_{j}$, where $b_{k j}$ is the transmission line susceptance between buses $k$ and $j$, and $V_k$ represents the voltage magnitude at bus $k$, which is assumed to be constant. The phase angle difference between buses $k$ and $j$ is denoted by $\delta_{k j}=\delta_{k}-\delta_{j}$.

The dynamics of the power system are further influenced by the turbine governor response, which introduces a delay in primary frequency control. This delay typically results in larger frequency deviations from the nominal grid frequency. The dynamics can be captured by the following first-order turbine governor model \cite{lee2019robustness}:
\begin{equation}
T_{k} \dot{p}_{k}+p_{k}+\frac{1}{R_{k}} \dot{\delta}_{k}=P_{G, k}, \quad k \in \mathcal{G}_G \label{eq: 1stO Turbin Gov Mdl}
\end{equation}
where $P_{G, k}$ is the reference power setpoint at generator bus $k$, $T_k$ represents the governor time constant, and $R_k$ is the speed droop coefficient.

To express the system model \eqref{eq: 2ndO Power Sys Dy} and \eqref{eq: 1stO Turbin Gov Mdl} in vector form, the following notation is defined. Let $\delta_{G}$ and $\delta_{L}$ be the vectors formed by stacking the scalar values $\delta_{k}$, for $k \in \mathcal{G}_G$, and $\delta_{k}$, for $k \in \mathcal{G}_L$, respectively. We then define $\delta=\begin{bmatrix}
    \delta_{G}^{\top} & \delta_{L}^{\top}
\end{bmatrix}^\top$. Similarly, let $p, P_{G}$, and $P_{L}$ be vectors that stack the scalar values $p_{k}, P_{G, k}$, for $k \in \mathcal{G}_G$, and $P_{L, k}$ for $k \in \mathcal{G}_L$, respectively. Additionally, let $M, D_{G}, D_{L}$, and $\Phi$ be diagonal matrices whose diagonal elements are $M_{k}$, $D_{k}$, for $k \in \mathcal{G}_G, D_{k}$, for $k \in \mathcal{G}_L$, and $\phi_{k j}$, for $(k, j) \in \mathcal{E}$, respectively. Lastly, let $E=\begin{bmatrix}
    E_{G}^{\top} & E_{L}^{\top}
\end{bmatrix}^\top$, where the subscripts $G$ and $L$ refer to the generator and load buses, respectively.

Now, let the disturbance vector be defined as $u=\begin{bmatrix}
    u_{G}^{\top} & u_{L}^{\top}
\end{bmatrix}^{\top}$. Using this, the system model equations \eqref{eq: 2ndO Power Sys Dy} and \eqref{eq: 1stO Turbin Gov Mdl} can be reformulated in a compact form as follows.
\begin{align}
M \ddot{\delta}_{G}+D_{G} \dot{\delta}_{G}+E_{G} \Phi \sin \left(E^{\top} \delta\right) & =p \nonumber\\
D_{L} \dot{\delta}_{L}+E_{L} \Phi \sin \left(E^{\top} \delta\right) & =P_{L}+u_{L}  \nonumber\\
T \dot{p}+p+R^{-1} \dot{\delta}_{G} & =P_{G}+u_{G}. \label{eq: Sys Mdl}
\end{align}

This straightforward formulation of disturbance encompasses a wide range of uncertainty scenarios, including load shedding, generation tripping, and unpredictable fluctuations in power output from wind turbines and solar panels. 

\subsection{Lur'e System Representation}
In the following, the system \eqref{eq: Sys Mdl} is reformulated as a Lur'e system, which is characterized by the interconnection of a linear dynamical system and a nonlinear static state feedback. 

Considering $u=\mathbf{0}$, let assume that $[{\delta_{G}^{*}}^{\top} \,\,\, {\delta_{L}^*}^{\top}]^\top$ and $\dot{\delta}=\mathbf{0}$ are the equilibrium point of \eqref{eq: Sys Mdl}, with a generator power injection of  $p^{*}$. The system state vector is then defined as $x=\begin{bmatrix} x_{1}^{\top} & x_{2}^{\top} & x_{3}^{\top} & x_{4}^{\top}\end{bmatrix}^{\top}$, where $x_{1}=\delta_{G}-\delta_{G}^{*}$, $x_{2}=\dot{\delta}_{G}$, $x_{3}=\delta_{L}-\delta_{L}^{*}$, and $x_{4}=p-p^{*}$.

Let $z=E^{\top} \delta-E^{\top} \delta^{*}$ represent the phase difference error on each transmission line, $y=\dot{\delta}_{G}$ represent the output vector containing the frequencies of the generators, and {$\varphi^{*}=E^{\top} \delta^{*}$}. With these new variables, the system \eqref{eq: Sys Mdl} can be expressed in the Lur'e form as follows:
\begin{subequations} \label{eq: Cmplt Mdl}
    \begin{align}
        \dot{x} & =A x+B_{v} v+B_{u} u \\
        v & =\sin \left(\varphi^{*}+z\right)-\sin \varphi^{*}-\operatorname{diag}\left(\cos \varphi^{*}\right) z \\
        y & =C_{y} x \\
        z & =C_{z} x,
    \end{align}
\end{subequations}
where 
\begin{align}
&A=  \begin{bmatrix}
\mathbf{0} & I & \mathbf{0} & \mathbf{0} \\
A_{21} & -M^{-1} D_{G} & A_{23} & M^{-1} \\
A_{31} & \mathbf{0} & A_{33} & \mathbf{0} \\
\mathbf{0} & -R^{-1} T^{-1} & \mathbf{0} & -T^{-1}
\end{bmatrix},    \nonumber\\
&B_v=\begin{bmatrix}
\mathbf{0} \\
-M^{-1} E_{G} \Phi \\
-D_{L}^{-1} E_{L} \Phi \\
\mathbf{0}
\end{bmatrix}, \ B_u= \begin{bmatrix}
\mathbf{0} & \mathbf{0} \\
\mathbf{0} & \mathbf{0} \\
\mathbf{0} & D_{L}^{-1} \\
T^{-1} & \mathbf{0}
\end{bmatrix}, \nonumber \\
&C_y=\begin{bmatrix} \mathbf{0} & I & \mathbf{0} & \mathbf{0}
        \end{bmatrix}, \ C_z=\begin{bmatrix} E_{G}^{\top} & \mathbf{0} & E_{L}^{\top} & \mathbf{0}
        \end{bmatrix} \label{eq: Lure form Sys Mdl}
\end{align}
with
\begin{align*}
& A_{21}=-M^{-1} E_{G} \Phi \operatorname{diag}\left(\cos \varphi^{*}\right) E_{G}^{\top} \\
& A_{23}=-M^{-1} E_{G} \Phi \operatorname{diag}\left(\cos \varphi^{*}\right) E_{L}^{\top} \\
& A_{31}=-D_{L}^{-1} E_{L} \Phi \operatorname{diag}\left(\cos \varphi^{*}\right) E_{G}^{\top} \\
& A_{33}=-D_{L}^{-1} E_{L} \Phi \operatorname{diag}\left(\cos \varphi^{*}\right) E_{L}^{\top}.
\end{align*}

The matrix $A$ in \eqref{eq: Lure form Sys Mdl} was derived by the linearization of the system described in \eqref{eq: Sys Mdl} around the equilibrium point $x^*=\mathbf{0}$. The vector $v$ in \eqref{eq: Cmplt Mdl} represents the static nonlinear feedback of the state $x$, specifically defined as $v=$ $\psi(z)=\psi\left(C_{z} x\right)$.

The model in \eqref{eq: Cmplt Mdl} has been discretized by the Euler method as follows \cite{butcher2016numerical}.
\begin{subequations} \label{eq: Dscrt Mdl}
    \begin{align}
        x_{t+1} & =(\Delta t A + I)  x_t + \Delta t B_{v} v_t+\Delta t B_{u} u_t \\
        v_t & =\sin \left(\varphi^{*}+z_t\right)-\sin \varphi^{*}-\operatorname{diag}\left(\cos \varphi^{*}\right) z_t \\
        y_t & =\begin{bmatrix} \mathbf{0} & I & \mathbf{0} & \mathbf{0}
        \end{bmatrix} x_t=C_{y} x_t \\
        z_t & =\begin{bmatrix} E_{G}^{\top} & \mathbf{0} & E_{L}^{\top} & \mathbf{0}
        \end{bmatrix} x_t=C_{z} x_t.
    \end{align}
\end{subequations}
where $\Delta t$ is a sampling step time for discretization. {It is important to choose a sufficiently small sampling time $\Delta t$ to accurately capture the system dynamics while ensuring the stability of the discretized model.}
\subsection{Frequency Response Requirements} \label{subsec: Mand Freq}
\begin{figure*}
\centering
\includegraphics[width=0.96\linewidth]{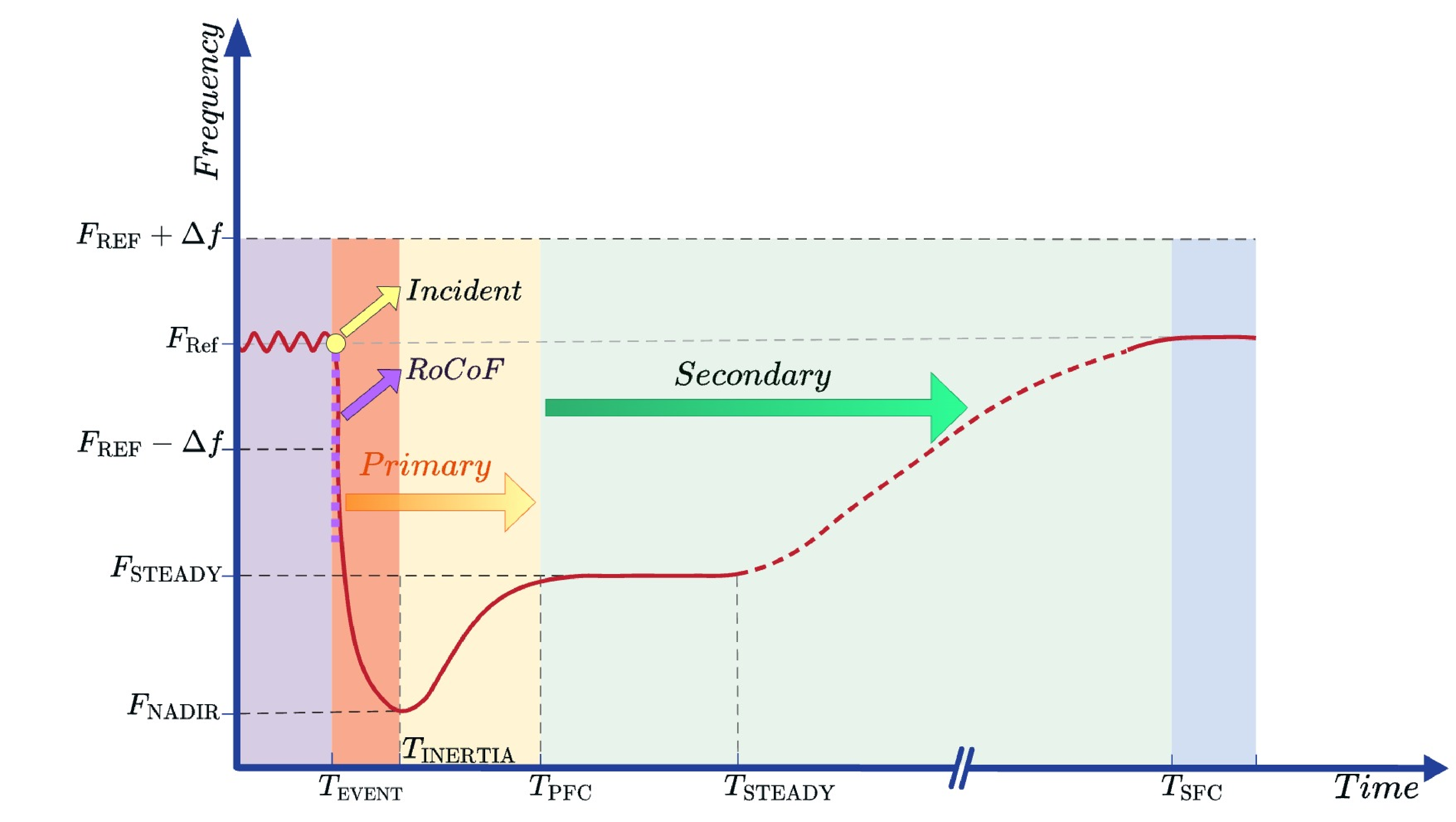}
    \caption{Frequency Response Evolution After a Contingency. Three phases can be seen in the frequency evolution: Inertial Response, Primary Frequency Control (PFC), and Secondary Frequency Control (SFC).} 
    \label{fig:freq resp}
\end{figure*}

Fig. \ref{fig:freq resp} illustrates the typical frequency behavior of a power system after a large loss of generation causes a frequency deviation \cite{liu2021inertia}. The frequency evolution can be categorized into three phases: Inertial Response, Primary Frequency Control (PFC), and Secondary Frequency Control (SFC). During the inertial response phase, the frequency dynamics are primarily influenced by the system's inertia, marked by a relatively high rate of frequency change known as the Rate of Change of Frequency (RoCoF). After this phase, the frequency progressively returns to its nominal value through the PFC and SFC stages \cite{kundur2007power}.
Power grids worldwide require automatic adjustments in active power output through frequency response services to maintain transient stability, though their specific parameters vary by region. For example, in the UK, the National Grid requires an automatic adjustment in active power output, known as Mandatory Frequency Response. This service aims to keep the frequency within statutory limits of $49.5 \, Hz$ to $50.5 \, Hz$, with stricter operational limits set between $49.8 \, Hz$ and $50.2 \, Hz$. The National Grid achieves this through three key response services: PFC, SFC, and High Frequency Response \cite{gridcode2023}.

Primary Response involves delivering additional active power or reducing demand within 10 seconds of a frequency event and sustaining this adjustment for 20 seconds. Secondary Response initiates a similar adjustment within 30 seconds and can maintain it for up to 30 minutes. In contrast, High-Frequency Response focuses on reducing active power within 10 seconds and sustaining this reduction indefinitely to maintain grid stability.

To represent the mandatory requirements for frequency response in a logical language, we aim to employ {STL}, which is explained in the subsequent section.
\subsection{STL Specifications}
An infinite run of the dynamical system can be considered as a signal {$\xi=x_0 \,x_1 \,x_2 \ldots$}, which is a sequence of observed states. We consider {STL} {formulae} with bounded-time temporal operators defined recursively according to the grammar \cite{oded2004monitoring}
\begin{equation*}
    \label{eq: stl}
    \varphi::= \top \,|\, \pi \,|\, \neg \varphi \,|\, \varphi_1 \wedge \varphi_2 \,|\, \varphi_1 \until_{[a,b]} \varphi_2 \,|\, \always\psi\,|\,\eventually\psi,
\end{equation*}
where $\pi$ is a predicate whose truth value is determined by the sign of a function, i.e., $\pi=$ $\{\alpha(x) \geq 0\}$ with $\alpha: \mathbb{R}^n \rightarrow \mathbb{R}$ being an affine function of state variables; $\varphi$, $\varphi_1$ and $\varphi_2$ are STL {formulae}; $\neg$ and $\wedge$ indicate negation and conjunction; and $\until_{[a, b]}$ is the until operator with $a, b \in \mathbb{R}_{\geq 0}$. A run $\xi$ satisfies $\varphi$ at time $k$, denoted by $(\xi, k) \models \varphi$, if the sequence {$x_k \, x_{k+1} \ldots$} satisfies $\varphi$. Accordingly, $\xi$ satisfies $\varphi$, if $(\xi, 0) \models \varphi$.

 Semantics of STL {formulae} are defined as follows. Every run satisfies $T$. The run $\xi$ satisfies $\neg \varphi$ if it does not satisfy $\varphi$; it satisfies $\varphi \wedge \psi$ if both $\varphi$ and $\psi$ hold. For a run {$\xi=x_0 \,x_1 \,x_2 \ldots$} and a predicate $\pi=\{\alpha(x) \geq 0\}$, we have $(\xi, k) \models \pi$ if {$\alpha(x_k) \geq 0$}. Finally, $(\xi, k) \models \bar{\varphi} \until_{[a, b]} \psi$ if $\varphi$ holds at every time step starting from time $k$ before $\psi$ holds, and additionally $\psi$ holds at some time instant between $a+k$ and $b+k$. Additionally, we derive the other standard operators as follows. Disjunction $\varphi \vee \psi:=\neg(\neg \varphi \wedge \neg \psi)$, the eventually operator as $\eventually_{[a, b]} \varphi:=\top \until_{[a, b]} \varphi$, and the always operator as $\square_{[a, b]} \varphi:=\neg \eventually_{[a, b]} \neg \varphi$. Thus $(\xi, t) \models \diamond_{[a, b]} \varphi$ if $\varphi$ holds at some time instant between $a+k$ and $b+k$ and $(\xi, k) \models \square_{[a, b]} \varphi$ if $\varphi$ holds at every time instant between $a+k$ and $b+k$.
\subsection{Robustness Semantics for STL Specification} \label{sec robust stl}
 In contrast to the above Boolean semantics, the quantitative semantics of STL. assigns to each formula $\varphi$ a real-valued function $\rho^{\varphi}$ of signal $\xi$ and $k$ such that $\rho^{\varphi}(\xi, k)>0$ implies $(\xi, k) \models \varphi$. Robustness of a formula $\varphi$ with respect to a run $\xi$ at time $k$ is defined recursively as
\begin{align*}
\rho^{\top}(\xi, k) & =+\infty \\
\rho^{\pi}(\xi, k) & ={\alpha(x_k)} \text { with } \pi=\{\alpha(x) \geq 0\} \\
\rho^{\neg \varphi}(\xi, k) & =-\rho^{\varphi}(\xi, k) \\
\rho^{\varphi \wedge \psi}(\xi, k) & =\min \left(\rho^{\varphi}(\xi, k), \rho^{\psi}(\xi, k)\right), \\
\rho^{\varphi \until_{[a, b]} \psi}(\xi, k) & = \\
& \hspace{-5mm}\max _{i \in[a, b]}\left(\min \left(\rho^{\psi}(\xi, k+i), \min _{j \in[0, i)} \rho^{\varphi}(\xi, k+j)\right)\right)
\end{align*}
where {$x_k$} refers to signal $\xi$ at time $k$. The robustness of the derived formula $\eventually_{[a, b]} \varphi$ can be worked out to be $\rho^{\eventually_{[a, b]} \varphi}(\xi, k)=$ $\max _{i \in[a, b]} \rho^{\varphi}(\xi, k+i)$; and similarly for $\square_{[a, b]} \varphi$ as $\rho^{\square_{[a, b]} \varphi}(\xi, k)$ $=\min _{i \in[a, b]} \rho^{\varphi}(\xi, k+i)$. The robustness of an arbitrary STL formula is computed recursively on the structure of the formula according to the above definition.
\section{PROBLEM FORMULATION} \label{sec: prob form}
In this section, we characterize the operational frequency requirements for the power system based on STL constraints. The STL framework allows us to precisely capture the essential frequency response characteristics that the power system must adhere to, such as the acceptable range of frequency deviations and the required settling time. 
Building upon the STL-based frequency constraints, we then formulate the research problem addressed in this paper.
\subsection{STL-based Frequency Constraints}
 Frequency requirements after a large power loss at a time point $T_{\textsc{EVENT}}$ can be formulated as temporal logic specifications as outlined in Section \ref{subsec: Mand Freq}. See Fig. \ref{fig:freq resp} for a graphical representation of the requirements and related parameters. Such a requirement is $y_t \models \Psi$, in which $\Psi  := \psi_0 \wedge \psi_1 \wedge \psi_2 \wedge \psi_3 \wedge \psi_4$ with
\begin{align}
    \psi_1 :=& \,\always_{[\frac{T_{\textsc{EVENT}}}{\Delta t},\frac{T_{\textsc{RoCoF}}}{\Delta t}]}(y_{t+1} - y_{t} \leq r), \nonumber\\
    \psi_2 :=& \,\always_{[\frac{T_{\textsc{EVENT}}}{\Delta t},\frac{T_{\textsc{Inertia}}}{\Delta t}]}(y_t \le F_{\textsc{Reference}}+\Delta f) \nonumber \\
    &\wedge (y_t \ge F_{\textsc{Nadir}})), \nonumber\\
    \psi_3 :=& \, \eventually_{[\frac{T_{\textsc{PFC}}}{\Delta t}, \frac{T_{\textsc{Steady}}}{\Delta t}]} \always_{[\frac{T_{\textsc{PFC}}}{\Delta t}, \frac{T_{\textsc{Steady}}}{\Delta t}]} ((y_t \ge F_{\textsc{Steady}}) \nonumber \\
    & \wedge (y_t \le F_{\textsc{Reference}}+\Delta f)), \nonumber\\
    \psi_4 := & \, \eventually_{[\frac{T_{\textsc{Steady}}}{\Delta t}, \frac{T_{\textsc{SFC}}}{\Delta t}]} \always_{[\frac{T_{\textsc{Steady}}}{\Delta t}, \frac{T_{\textsc{SFC}}}{\Delta t}]} ((y_t \ge F_{\textsc{REF}}-\Delta f) \nonumber \\
    & \wedge (y_t \le F_{\textsc{REF}}+\Delta f)).  \label{eq dscrt stl}
\end{align}
\subsection{Problem Statement}
Consider the power system \eqref{eq: 2ndO Power Sys Dy} written in the Lur'e form \eqref{eq: Lure form Sys Mdl}, with initial condition $x_{0}=\mathbf{0}$.
We redefined the disturbance $u_t$ in the power system model \eqref{eq: Cmplt Mdl} as $u_t = \mu w_t$, where $w_t$ is { the normalized disturbance and $\mu$ is the normalization factor.}

 The objective of our problem is to find the maximum $\mu$ such that for all $w_t$ vectors with $\|w_t\|_{\infty} \le 1$ and $t\in \mathbb{N}_0$, the following two conditions hold:
\begin{align}
    &\|z_t\|_{\infty} \leq \bar{z}, \quad \forall t \in \mathbb{N}_0; \label{cond z}\\
    &\rho^\Psi(y,t) \ge \epsilon,
    \quad\forall t \in \mathbb{N}_0,
    \label{cond y}
\end{align}
where $\rho^\Psi(y,t)$ is the robustness $y =y_0 \,y_1 \,y_2 \ldots$ against the specification $\Psi$. 

Condition \eqref{cond z} introduces a constraint $\bar{z}$ on the angular difference between adjacent buses, preventing the generators from losing synchronization during transient dynamics. Condition \eqref{cond y} ensures that the frequency requirements based on grid operational constraints, such as those imposed by the UK National Grid \cite{gridcode2023}, are met with minimum robustness threshold $\epsilon \ge 0$. Considering these conditions, we can define the main problem we aim to address.
\begin{resp1}
\begin{problem}
\label{prblm}
    Given the system model in \eqref{eq: Dscrt Mdl}, calculate the maximum {normalization factor of disturbance,} {$\mu$}, on the {normalized} disturbance{, $w_t$,} s.t. conditions \eqref{cond z} and \eqref{cond y} holds for all $w_t$ vectors with $\|w_t\|_\infty \le 1$. \label{prblm: max bound dist}
\end{problem}
\end{resp1}
%
%
\section{COMPUTATION OF LARGEST DISTURBANCE} \label{sec: large dist}
In order to solve Problem \ref{prblm: max bound dist}, the following optimization problem is defined:
\begin{align}
\max_{\mu} \,\,\, &\mu \nonumber \\
\st\quad& \text{for all } w_t \text{ with } \|w_t\|_\infty\le 1 \text{ and all } t\in\mathbb N_0:\nonumber\\  
& \rho^\Psi(y,0) \ge \epsilon, \text{ with } y = (y_0,y_1,y_2,\ldots) ; \nonumber \\
 &x_{t+1} =(\Delta t A + I)  x_t + \Delta t B_{v} v_t+\Delta t B_{u}\mu  w_t; \nonumber\\
        &v_t =\sin \left(\varphi^{*}+z_t\right)-\sin \varphi^{*}-\operatorname{diag}\left(\cos \varphi^{*}\right) z_t;  \nonumber\\
        &y_t =C_{y} x_t;  \nonumber\\
        & z_t =C_{z} x_t;  \nonumber \\
        & \|z_t\|_\infty\le \bar{z}. 
 \label{eq opt 1}
 \end{align}

In the proposed optimization in \eqref{eq opt 1}, we aim to maximize $\mu$ for all $w_t$ satisfying $ \|w_t\|_\infty\le 1,\,\forall t\in\mathbb N_0$. Hence, we will use robust nonconvex scenario optimization framework proposed in \cite{garatti2024non}, which is discussed in next section.
\subsection{Robust nonconvex Scenario Optimization}
The optimization problem in \eqref{eq opt 1} can be equivalently expressed in scenario optimization form. To do this, consider uniform distribution on the space $\mathcal{W} = [-1,1]^{m+n}$ and get $Q$ sampled scenarios for the disturbance as $\{w^1,w^2,\cdots,w^Q\} \in \mathcal{W}$ with samples $w^i = (w_0^i,w_1^i,\cdots,w_{N-1}^i)$. Then, the optimization problem in \eqref{eq opt 1} can be approximated as follows:
\begin{align}
\max_{\mu} \,\,\, &\mu \nonumber \\
\st\quad& \text{for all } i\in\{1,2,\ldots,Q\} \text{ and all } t\in\mathbb N_0:\nonumber\\  
& \rho^\Psi(y^i,0) \ge \epsilon,\text{ with } y^i = (y_0^i,y_1^i,y_2^i,\ldots);  \nonumber \\
 &x_{t+1}^i =(\Delta t A + I)  x_t^i + \Delta t B_{v} v_t^i+\Delta t B_{u}\mu  w^i, \nonumber \\
 & \quad\text{ with } w^i = (w_0^i,w_1^i,\cdots,w_{N-1}^i);\nonumber\\
        &v_t^i =\sin \left(\varphi^{*}+z_t^i\right)-\sin \varphi^{*}-\operatorname{diag}\left(\cos \varphi^{*}\right) z_t^i;  \nonumber\\
        &y_t^i =C_{y} x_t^i;  \nonumber\\
        & z_t^i =C_{z} x_t^i; \nonumber \\
 & \|z_t^i\|_\infty\le \bar{z}.  
 \label{eq scenario opt 1}
 \end{align}
 
To rigorously evaluate the robustness of our solution, we incorporate risk and confidence analysis within a scenario optimization framework proposed in \cite{garatti2024non}. In this context, the \textbf{risk} \( V(\mu^*_Q) \) represents the probability that the optimal solution of \eqref{eq scenario opt 1}, denoted as \( \mu^*_Q \), will fail to meet the requirements of new unseen scenarios thus being an infeasible solution for \eqref{eq opt 1}. This risk is assessed through the \textbf{support complexity} \( s^*_Q \), which identifies the smallest subset of scenarios that are critical for determining \( \mu^*_Q \).

To ensure confidence in this risk assessment, we apply Theorem \ref{thrm conf} stated below, which defines a threshold function \( \varepsilon(s^*_Q) \) to achieve a confidence level of \( (1 - \beta) \).

\begin{theorem}\textbf{(\cite{garatti2024non}, Thm. 6)}
\label{thrm conf}
Given $\beta \in(0,1)$, for any $k=0,1, \ldots, Q-1$, the polynomial equation in the $\iota$ variable
\begin{equation*}
    \frac{\beta}{Q+1} \sum_{q=k}^Q\binom{q}{k} \iota^{q-k}-\binom{Q}{k} \iota^{Q-k}=0
\end{equation*}
has one and only one solution $\iota(k)$ in the interval $(0,1)$. Letting $\varepsilon(k):=1-\iota(k), k=$ $0,1, \ldots, Q-1$, and $\varepsilon(Q)=1$, it holds that
\begin{equation*}
    \mathbb{P}^{Q}\left\{V\left(\mu_{Q}^{*}\right)>\varepsilon\left(s_{Q}^{*}\right)\right\} \leq \beta . 
\end{equation*}
\end{theorem}
\smallskip

This theorem establishes that for any confidence level \( \beta \) within the interval \( (0, 1) \), there exists a unique solution \( \iota(k) \) for each complexity level \( k \), allowing us to set \( \varepsilon(k) = 1 - \iota(k) \). Thus, Theorem \ref{thrm conf} provides a probabilistic upper bound, ensuring that the likelihood of \( V(\mu^*_Q) \) exceeding \( \varepsilon(s^*_Q) \) remains within the desired confidence level, which enables us to confidently assess the robustness of our solution against unforeseen scenarios.

After addressing uncertainty in the scenario optimization \eqref{eq scenario opt 1}, next subsection discusses how to manage STL robustness specifications in the optimization.
{\begin{remark}[Feasiblity of the Optimization Problem \eqref{eq scenario opt 1}]
Although Problem \eqref{eq scenario opt 1} is a nonlinear optimization problem and may not always guarantee global optimality, it is important to note that, in the context of power systems, which are typically designed to meet frequency requirements under normal operating conditions, a feasible solution always exists. Specifically, in the absence of disturbances (when $\mu = 0$), the system operates nominally and satisfies the STL specifications. The objective of the optimization is to determine the \emph{largest disturbance magnitude} that can be tolerated while still maintaining these properties. In the worst-case scenario, the solution to \eqref{eq scenario opt 1} may indicate a disturbance bound of zero. This means that the system cannot accommodate any deviation while still satisfying frequency requirements. Therefore, the formulation is always feasible when the system satisfies the requirements under no disturbance, and the value of $\mu$ provides valuable information about the system’s resilience, regardless of the nonconvex nature of the problem.
\end{remark}}
\subsection{Mixed Integer Nonlinear Encoding of STL} \label{sec minlp}

The STL robustness in the optimization \eqref{eq scenario opt 1} can be represented as a Mixed Integer Nonlinear Program (MINLP). Following the robustness-based encoding approach outlined in \Negar{\cite{raman2015reactive, farahani2018formal}}, each STL formula \(\Psi\) can be transformed into a set of mixed-integer constraints. This method encodes the system trajectory as a finite sequence of states that adhere to the model's dynamics, with the STL formula \(\Psi\) expressed through a series of mixed-integer constraints.

For each sub-formula of \(\Psi\), a robustness variable \(\rho_{k}^{\Psi}\) is introduced. Constraints are then applied such that \(\rho_{k}^{\Psi} > \epsilon\) indicates that the formula \(\Psi\) is satisfied at time \(k\). By utilizing this encoding, operations like maximum and minimum are managed through additional binary variables and a large constant (often referred to as big-M), which allows the nonconvex STL constraints to be incorporated into the optimization framework. This formulation effectively integrates STL constraints, ensuring that the generated system trajectory complies with both the system dynamics and the STL specifications.

As a result, we propose the following algorithm to address Problem \ref{prblm}.
\RestyleAlgo{ruled}
\IncMargin{0.1em}
\begin{algorithm}
\hspace{3mm} 
\begin{minipage}{\dimexpr\linewidth-3mm}
\SetAlgoLined
\SetKwInOut{Input}{Input}
\SetKwInOut{Output}{Output}
\SetKwFunction{GenerateRandomNumber}{GenerateRandomNumber}
\SetKwFunction{Wait}{Wait}
\SetAlgoNlRelativeSize{-1}
\caption{STL-Based Resilience Assessment Algorithm} 
\label{alg res assess}
\setcounter{AlgoLine}{0}
\BlankLine
\Input{Matrices $M$, $D_G$, $E_G$, $D_L$, $E_L$, $T$, $R$, vectors $P_L$, $P_G$ and $p$ for power system model, parameters $T_{\textsc{Event}}$, $T_{\textsc{RoCoF}}$, $T_{\textsc{Inertia}}$, $T_{\textsc{PFC}}$, $T_{\textsc{STD}}$, $T_\textsc{SFC}$, $\textsc{RoCoF}_{\max}$, $F_{\textsc{Nadir}}$, $F_{\textsc{Steady}}$, $F_{\textsc{Ref}}$, and $\Delta f$ for frequency responce requirement and confidence parameter $\beta$}

Define power system model as \eqref{eq: Sys Mdl}

Represent the power system model as Lur'e system in \eqref{eq: Cmplt Mdl}

Discretisize Lur'e system using the Euler method as in \eqref{eq: Dscrt Mdl}

Determine STL specifications $\Psi$ based on frequency response requirements in \eqref{eq dscrt stl}

Define robustness semantics $\rho^{\Psi}(\xi, k)$ according to Section \ref{sec robust stl}

Define optimization problem as \eqref{eq opt 1}

Based on Theorems \ref{thrm conf}, and the parameter $\beta$, decide on the number of scenarios for variable $w_t$, denoted $Q$, in a manner that ensures $\varepsilon$ remains small

Redefine optimization problem in scenario programming framework as \eqref{eq scenario opt 1}

Transform STL semantics in \eqref{eq scenario opt 1} using the Big M method, encode the optimization \eqref{eq scenario opt 1} to MINLP as described in \ref{sec minlp}

Solve the optimization problem using MINLP solvers

\Output{{\emph{Maximum} $\mu^\ast_Q$}}
\BlankLine
\end{minipage}
\end{algorithm}
\begin{remark}[Selecting $Q$ in Algorithm~\ref{alg res assess}]
The parameter \( Q \) in Step 7 of the Algorithm~\ref{alg res assess} determines the number of scenarios sampled for the scenario optimization problem \eqref{eq scenario opt 1} and directly influences the probabilistic guarantees derived from Theorem~\ref{thrm conf}. There is a trade-off between the number of samples \( Q \), the confidence level \( \beta \), and the risk parameter \( \epsilon \). According to the scenario optimization framework, for a fixed confidence level \( \beta \), increasing \( Q \) decreases the risk \( \epsilon \), which in turn enhances the probabilistic guarantee regarding the solution's feasibility under unseen disturbances. However, increasing \( Q \) can also lead to higher computational costs. In practice, we fix \( \beta \) and adjust the value of \( Q \) to empirically achieve the desired balance between risk, confidence, and computation costs.
\end{remark}
\begin{remark}
    In order to implement Algorithm \ref{alg res assess}, we use Pyomo package \cite{bynum2021pyomo, hart2011pyomo}. To solve the final optimization problem in Step 10 of the algorithm, we use the Mixed-Integer Nonlinear Decomposition Toolbox in Pyomo (MindtPy) solver.
\end{remark}
\section{SIMULATION RESULTS} \label{sec: sim res}
In this section, we apply the proposed approach to the Single Machine Infinite Bus system, as described in \cite{lee2019robustness} and IEEE 9-Bus Benchmark system adopted from \cite{zim2011matpower}. We define the STL specifications for each case study according to the UK frequency requirements using \eqref{eq dscrt stl} {to $\frac{y}{2\pi}$}. A summary of the parameters for the mandatory frequency response is provided based on the UK National Grid Code in Table \ref{tab:parameters} \cite{gridcode2023}.
\begin{table}[h!]
    \centering
    \caption{Parameters of the Mandatory Frequency Response for the UK Power Grid Case Study \eqref{eq dscrt stl}}
    \begin{tabular}{|c||c||c||c||c||c|}
        \hline
        \textbf{Parameter} & \textbf{Value} & \textbf{Unit} & \textbf{Parameter} & \textbf{Value} & \textbf{Unit}\\ \hline
            $T_{\textsc{Event}}$    & 0           & s      &  $\textsc{RoCoF}_{\max}$  & $ 1$  & Hz/s\\ \hline
        $T_{\textsc{RoCoF}}$    & 500         & ms    & $F_{\textsc{Nadir}}$    & 49.2        & Hz \\ \hline
        $T_{\textsc{Inertia}}$  & 10          & s      & $F_{\textsc{Steady}}$   & 49.5        & Hz  \\ \hline
        $T_{\textsc{PFC}}$  & 30          & s   & $F_{\textsc{Ref}}$ & 50          & Hz      \\ \hline
        $T_{\textsc{STD}}$   & 60          & s      & $\Delta f$                & 0.2         & Hz  \\ \hline
        $T_\textsc{SFC}$                   & 30          & min   &&& \\\hline  
    \end{tabular}
    \label{tab:parameters}
\end{table}
The simulation case studies are conducted for the deviation of frequency from the nominal frequency of $F_{\textsc{REF}}=50$~Hz. In all of the case studies, we used the Pyomo package and MindtPy in the PyCharm Python IDE to implement Algorithm \ref{alg res assess}. After calculating the maximum $\mu^*_Q$, we plotted the simulation results in MATLAB. The computations are performed on a MacBook Pro with the M2 Pro chip and 16GB of memory.

Since the signal $w_t$ is regarded as unknown in our analysis, and given the focus on verifying the largest disturbances, we lack knowledge on how to handle disturbance incidents within the power grid. However, based on the concepts of PFC and SFC stages discussed in Section \ref{subsec: Mand Freq}, we have identified two realistic scenarios for the power systems presented in the scenarios described below.
\begin{itemize}
    \item \textbf{\emph{Scenario 1:}} At \( t = 0^- \), the generation and load are assumed to be balanced, and the system frequency is at its nominal value. At \( t = 0^+ \), an unexpected loss of generation (generator tripping) occurs, causing the frequency to begin decreasing at a relatively high rate. We model it as the largest power loss by assuming that $w_t = -1$ at the beginning of the incident. We assume that, over time, power injection and withdrawal will balance out. Synchronous generator turbine-governors and inverter-based resource active power-frequency control systems adjust their active power output in response to frequency deviations. We represent this as a decrease of a certain percentage of \(\|w_t\|_\infty\) in two stages over time until \(w_t\) reaches \(0\).
    \item \textbf{\emph{Scenario 2:}}
    At \( t = 0^- \), the generation and load are balanced, and the system frequency is at its nominal value. At \( t = 0^+ \), an unexpected loss of generation occurs, causing the frequency to begin decreasing at a relatively high rate. We model it as the largest power loss by assuming that $w_t$ is a uniform distribution over $[-1,0]$ at the beginning of the incident. We assume that, over time, power injection and withdrawal will balance out. Synchronous generator turbine-governors and inverter-based resource active power-frequency control systems adjust their active power output in response to frequency deviations. We represent this as a reduction of a certain percentage of uniform distribution of \(\|w_t\|_\infty\) in several stages over time. Then, we assume that $w_t$ follows a uniform distribution over the interval $[0,1]$, representing active power injection to compensate for the initial power loss for some period of time.
\end{itemize}

\subsection{Single Machine Infinite Bus }
Consider a system composed of a single machine connected to an infinite bus through a lossless transmission line. The dynamic equation is given by
\begin{equation*}
M \ddot{\delta}+D \dot{\delta}+\phi \sin \delta=p+u,
\end{equation*}
where parameters defined as \( M = 1 \), \( D = 1.2 \), \( p = 0.2 \), and \( \phi = 0.8 \). For \( u = 0 \), the equilibrium is represented by \( \delta^{*} = \arcsin\left(\frac{p}{\phi}\right) \) with \( \dot{\delta} = 0 \). The output, expressed as frequency in Hz, is given by \( y = \frac{\dot{\delta}}{2 \pi} \). By substituting \( z = \delta - \delta^{*} \) and \( v = \sin(\delta) - \cos(\delta^{*}) w \), we obtain the following results
\begin{equation}
M \ddot{x}+D \dot{x}+\phi \cos \left(\delta^{*}\right) x+\phi v=u. \label{eq sngl mch inf bus}
\end{equation}

\begin{figure}[tbp]
    \centering
    \subfloat[]{\includegraphics[width=0.49\linewidth]{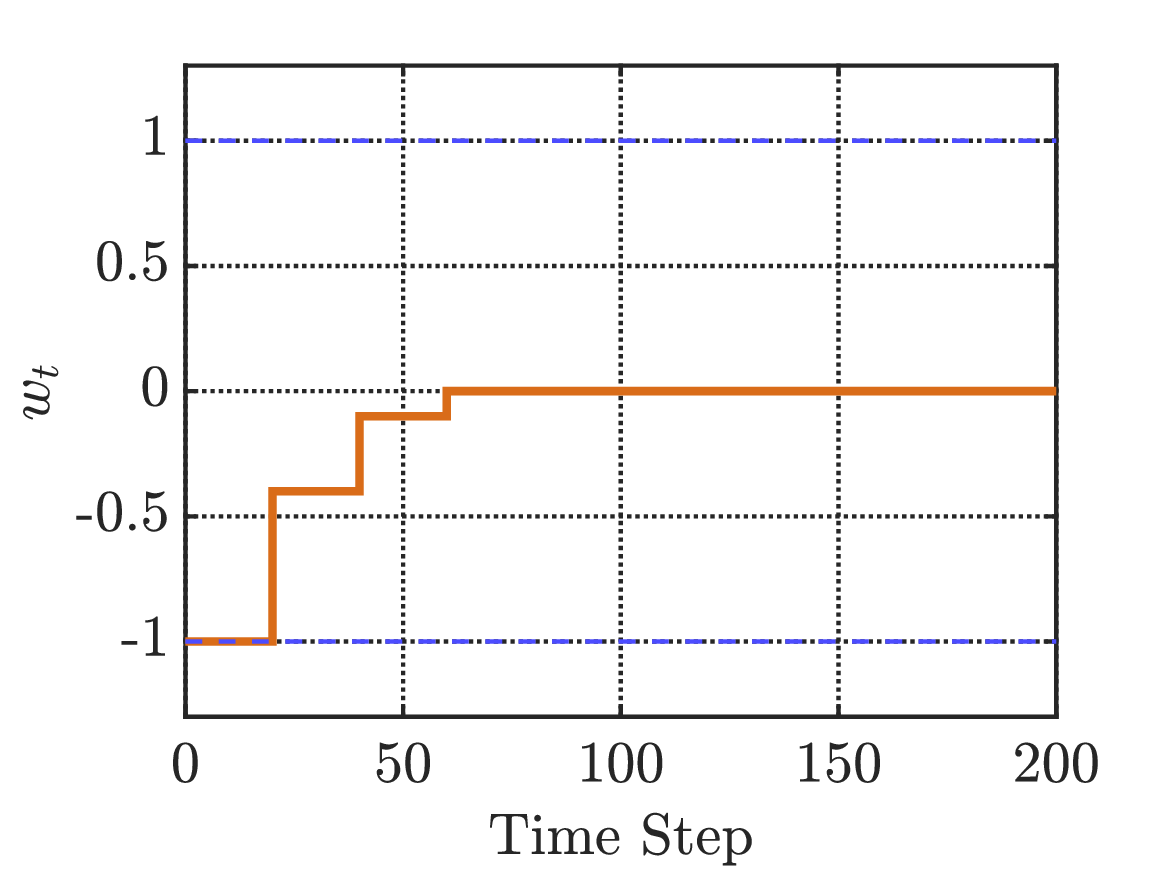} }
    \hfill
    \subfloat[]{\includegraphics[width=0.49\linewidth]{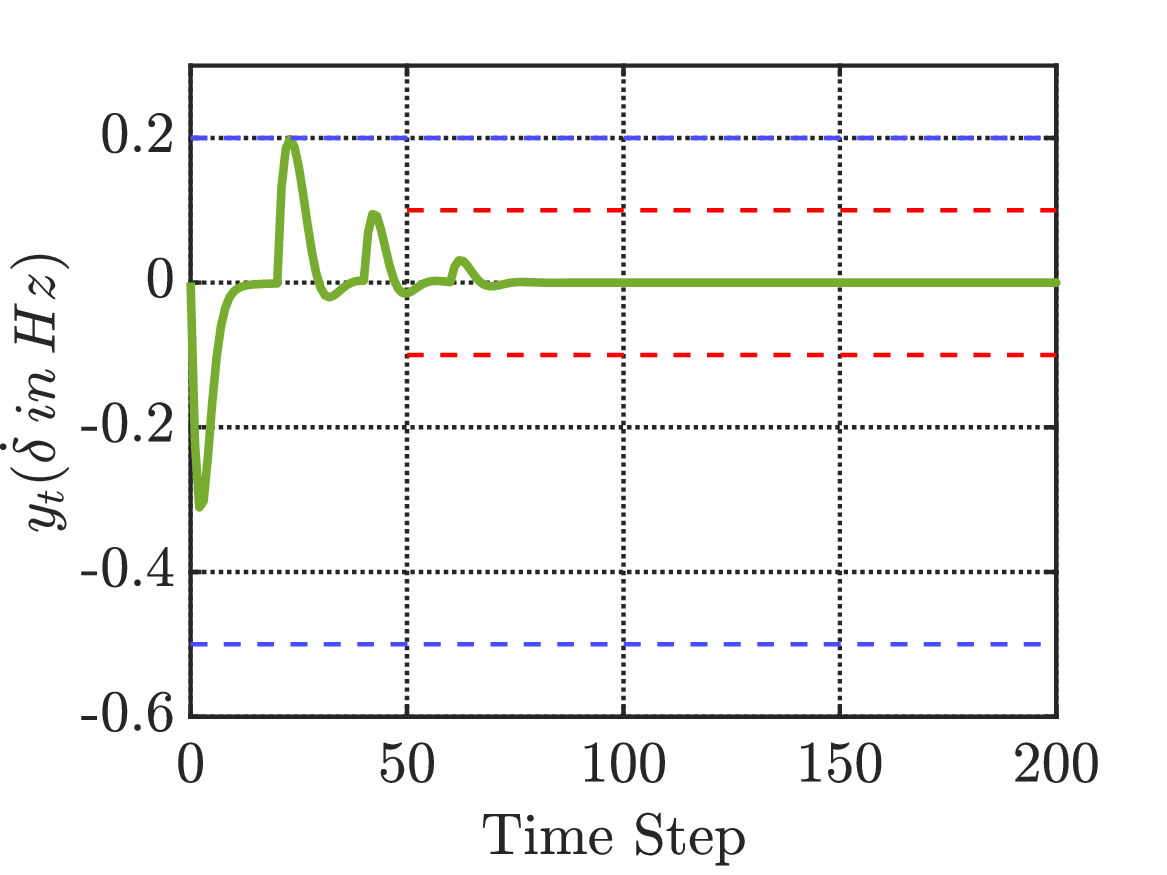}} \\
    \subfloat[]{\includegraphics[width=0.49\linewidth]{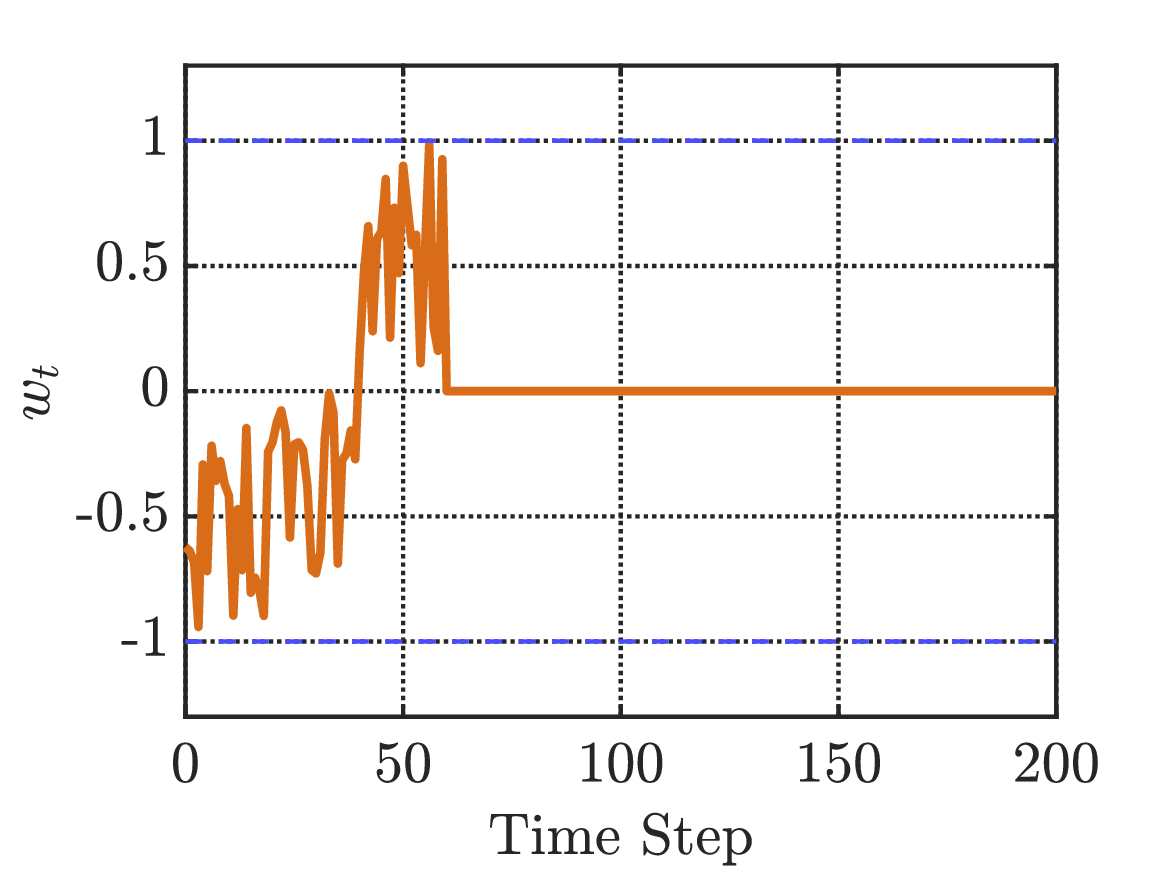}}
    \hfill
    \subfloat[]{\includegraphics[width=0.49\linewidth]{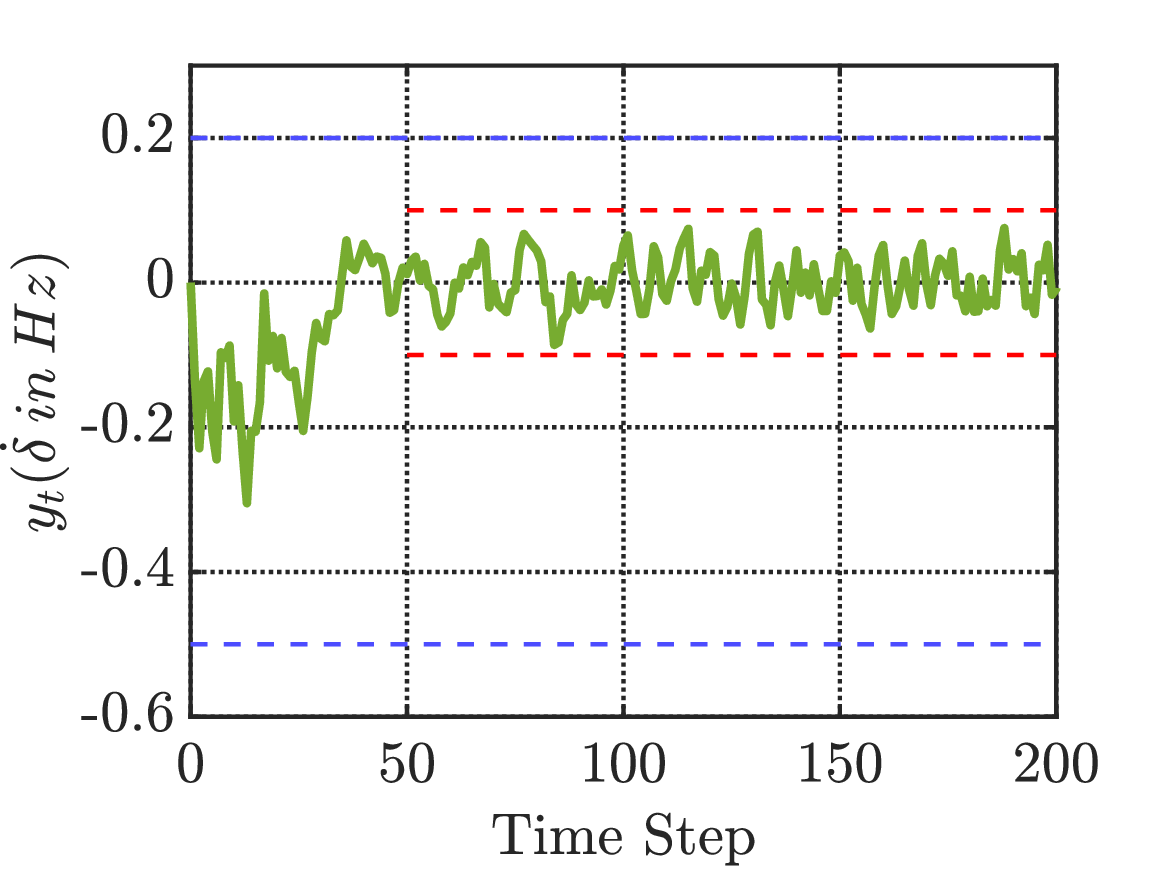}}
    \caption{Frequency response of single machine infinite bus case study by applying two scenarios. (a) $w_t$ designed based on \emph{Scenario 1} and (b) frequency response in \emph{Scenario 1}. (c) $w_t$ designed based on \emph{Scenario 2} and (d) frequency response in \emph{Scenario 2}.}
    \label{fig:sngl mchn}
\end{figure}

Using $\beta = 0.05$ and a discretization time of $\Delta t = 0.5$~s, Algorithm \ref{alg res assess} was applied to system \eqref{eq sngl mch inf bus}. The algorithm ran for $23.6891$ seconds, and the maximum value of $\mu$ was calculated as $0.7746$ p.u. This maximum disturbance is larger than the maximum disturbance calculated in \cite{lee2019robustness}, which was found to be less than $0.6$. Simulation results for the two mentioned scenarios are illustrated in Fig \ref{fig:sngl mchn}. The results show that the frequency constraints in \eqref{eq dscrt stl} with the grid code requirements given in Table~\ref{tab:parameters} are met under these two scenarios.
\subsection{IEEE 9-Bus Benchmark System}
This section presents numerical case studies of the IEEE 9-Bus System. Using a value of $\beta = 0.05$ and a discretization time of $\Delta t = 0.005$~s, Algorithm \ref{alg res assess} was applied to the 9-bus system, whose details are given in \cite{IEEE9bus}. The algorithm was executed over a period of $266.8387$ seconds, and the maximum value of $\mu$ was computed to be $1.6438$ p.u. In this study, we applied disturbances to buses 1, 3, 7, and 9, considering both scenarios for each bus. The simulation results are shown in Fig \ref{fig:sngl mchn}. The $j^{th}$ element of $y_t$ and $w_t$ are shown by notation of $y[j]$ and $w[j]$ in the sub-figures in Fig~\ref{fig:sngl mchn}. {The results in this figure indicate that the frequency constraints in \eqref{eq dscrt stl} are satisfied in both scenarios.}
\begin{figure}[htbp]
    \centering
    \subfloat[]{\includegraphics[width=0.49\linewidth]{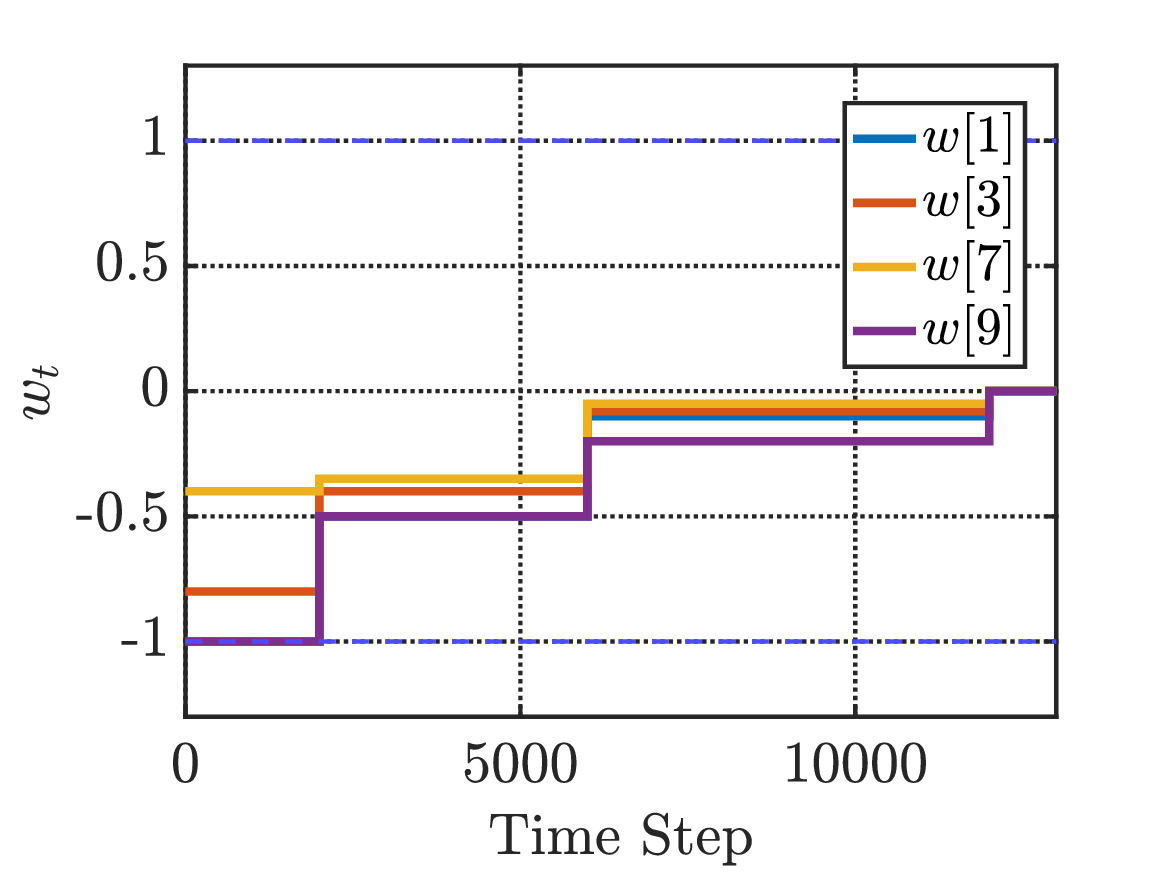} }
    \hfill
    \subfloat[]{\includegraphics[width=0.49\linewidth]{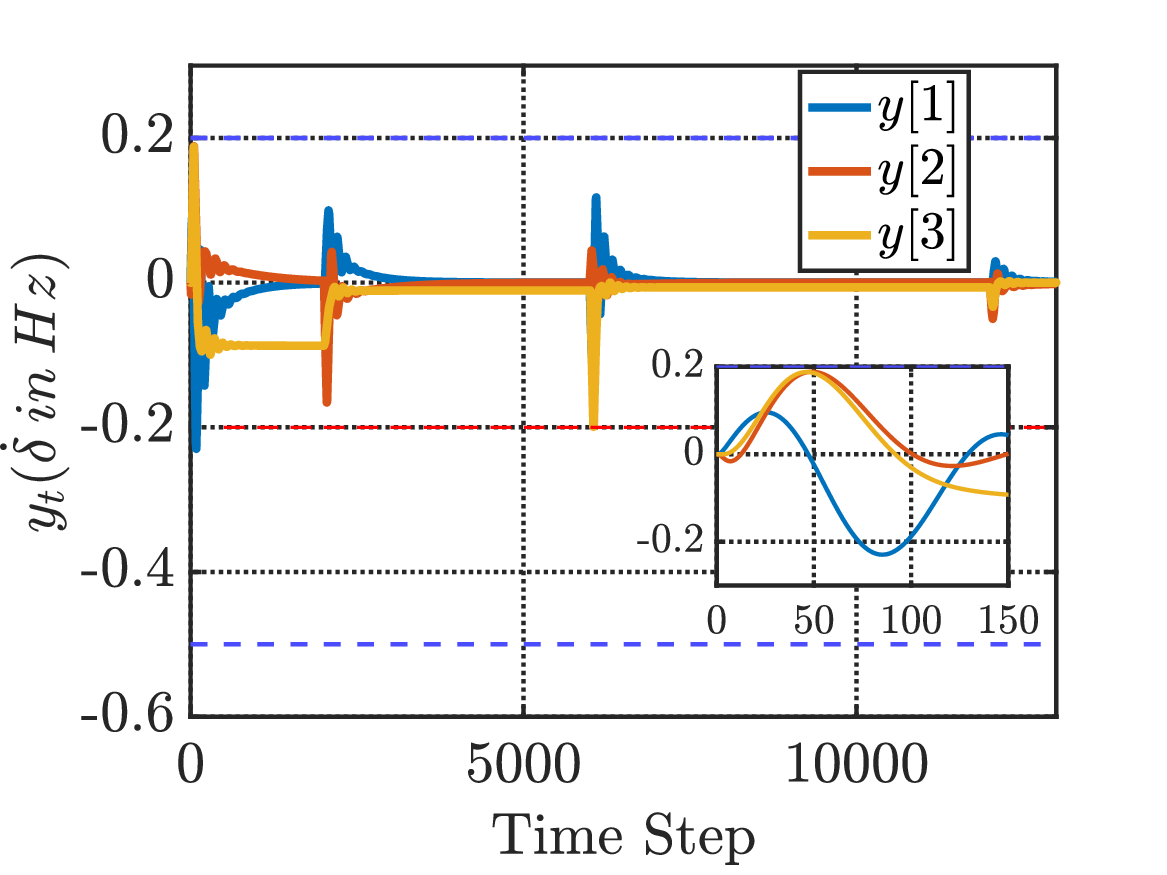}} \\
    \subfloat[]{\includegraphics[width=0.49\linewidth]{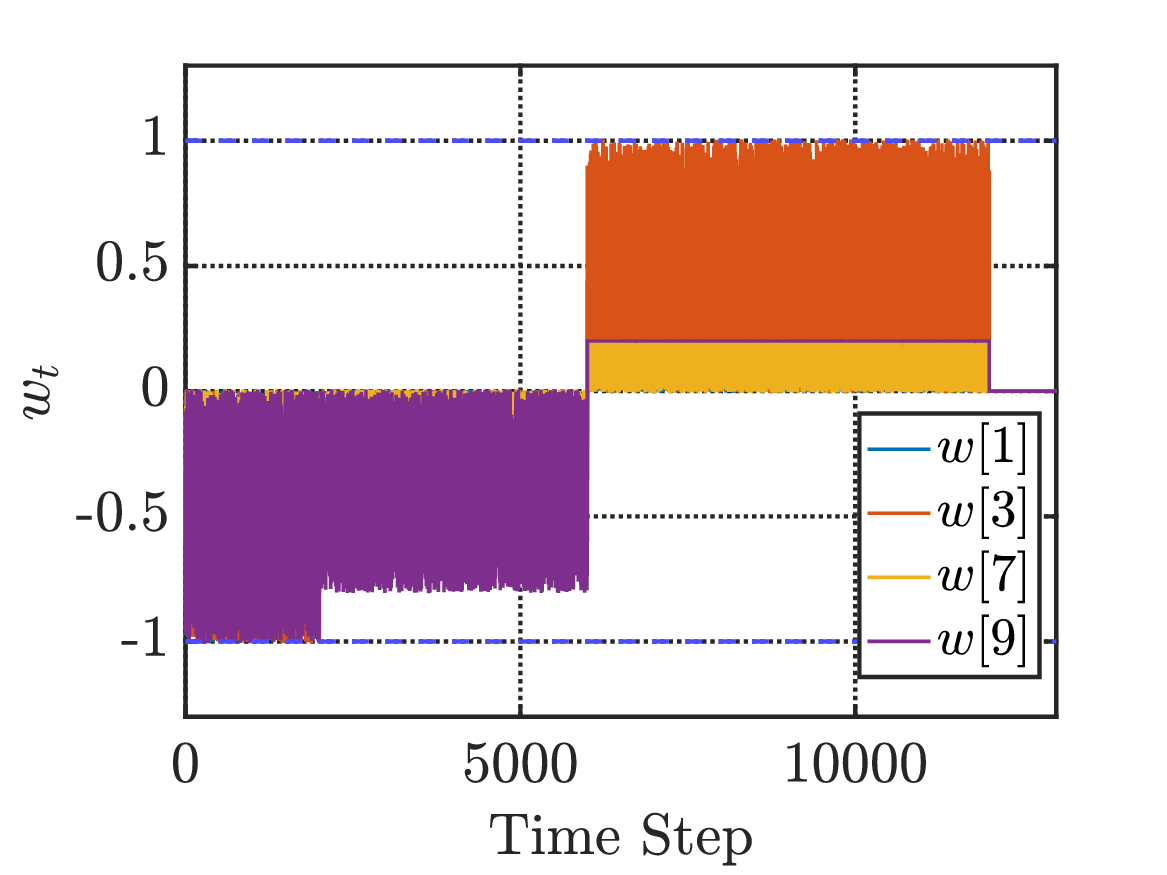}}
    \hfill
    \subfloat[]{\includegraphics[width=0.49\linewidth]{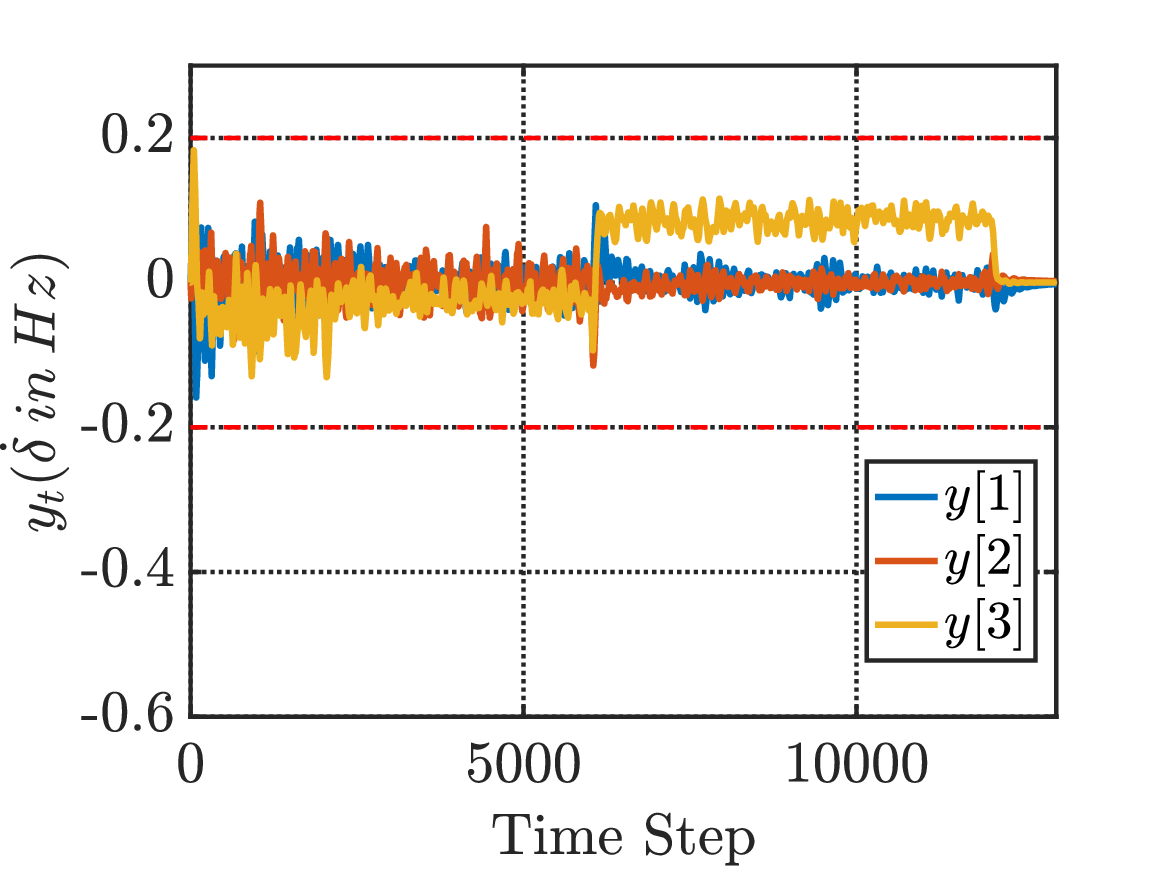}}
    \caption{Frequency response of IEEE 9-Bus system by applying two scenarios. (a) $w_t$ designed based on \emph{Scenario~1} for buses 1, 3, 7, and 9, and (b) frequency response in \emph{Scenario~1}. (c) $w_t$ designed based on \emph{Scenario 2} for buses 1, 3, 7, and 9 and (d) frequency response in \emph{Scenario 2}.}
    \label{fig:nine bus}
\end{figure}

\section{CONCLUSIONS}
\label{sec: concl}
This paper presents a framework for assessing the resilience of power systems to disturbances under STL-based frequency constraints. By utilizing the Lur’e system representation to model power system dynamics, we capture the essential nonlinear characteristics that affect the system evolution. The introduction of an STL robustness metric allows for precise frequency constraints that align with grid operator requirements, thereby enhancing the reliability of transient evaluations. We model the calculation of the largest disturbance that the power system can handle, based on frequency response requirements, as an optimization problem with STL robustness constraints, which is {an uncertain nonconvex optimization problem}. To address the uncertainties in the proposed optimization problem, we utilized a scenario optimization framework and translated the optimization problem into a nonconvex scenario optimization problem. {In order to solve the optimization problem, the STL constraints are encoded in a MINLP setting.} By solving this optimization problem, we were able to determine the maximum level of disturbance. To illustrate the effectiveness of our proposed algorithm, it is applied to the Single Machine Infinite Bus case study and the IEEE 9-Bus system, where we obtained promising results. Unlike the previous method proposed in \cite{lee2017robust}, our algorithm incorporates more realistic frequency constraints, resulting in more accurate outcomes. However, this improvement in accuracy has led to an increase in computation time. In our future work, we plan to analyze the resilience of various connection topologies in relation to power losses and extend our work to inverter-dominated power systems. Additionally, we aim to incorporate different scenarios for the normalization vector \( w \) into the optimization problem. This will allow us to calculate disturbances based on a semi-known \( w \) vector and achieve more realistic results.
\bibliographystyle{IEEEtran}
\bibliography{Ref}
\end{document}